\documentclass[aps,prc,twocolumn,superscriptaddress]{revtex4-2}
\usepackage{graphicx}
\usepackage{hyperref}

\begin{document}


\title{Blinding for precision scattering experiments: The MUSE approach as a case study}

\author{J.~C.~Bernauer}
\email{jan.bernauer@stonybrook.edu}
\affiliation{Center for Frontiers in Nuclear Science, Stony Brook University, Stony Brook, NY 11794, USA}
\affiliation{RIKEN BNL Research Center, Brookhaven National Laboratory, Upton, NY 11973, USA}
\author{E.~W.~Cline}
\affiliation{Center for Frontiers in Nuclear Science, Stony Brook University, Stony Brook, NY 11794, USA}
\affiliation{Laboratory for Nuclear Science, Massachusetts Institute of Technology, Cambridge, MA 02139, USA}
\author{H.~Atac}
\affiliation{Department of Physics, Temple University, Philadelphia, PA 19122, USA}
\author{W.~J.~Briscoe}
\affiliation{Department of Physics, The George Washington University, Washington, D.C. 20052, USA}
\author{A.~Christopher~Ndukwe}
\affiliation{Department of Physics, Hampton University, Hampton, VA 23668, USA}
\author{E.~J.~Downie}
\affiliation{Department of Physics, The George Washington University, Washington, D.C. 20052, USA}
\author{I.~P.~Fernando}
\affiliation{Department of Physics, Hampton University, Hampton, VA 23668, USA}
\author{T.~Gautam}
\affiliation{Department of Physics, Hampton University, Hampton, VA 23668, USA}
\author{R.~Gilman}
\affiliation{Rutgers, The State University of New Jersey, Piscataway, NJ 08855, USA}
\author{R.~Goldin}
\affiliation{Racah Institute of Physics, The Hebrew University of Jerusalem, Jerusalem, Israel}
\author{M.~Kohl}
\affiliation{Department of Physics, Hampton University, Hampton, VA 23668, USA}
\author{I.~Lavrukhin}
\affiliation{Department of Physics, The George Washington University, Washington, D.C. 20052, USA}
\affiliation{Randall Laboratory of Physics, University of Michigan, Ann Arbor, Michigan 48109, USA}
\author{W.~Lin}
\affiliation{Rutgers, The State University of New Jersey, Piscataway, NJ 08855, USA}
\author{W.~Lorenzon}
\affiliation{Randall Laboratory of Physics, University of Michigan, Ann Arbor, Michigan 48109, USA}
\author{P.~Mohanmurthy}
\affiliation{Laboratory for Nuclear Science, Massachusetts Institute of Technology, Cambridge, MA 02139, USA}
\author{S.~J.~Nazeer}
\affiliation{Department of Physics, Hampton University, Hampton, VA 23668, USA}
\author{M.~Nicol}
\affiliation{Department of Physics and Astronomy, University of South Carolina, Columbia, SC 29208, USA}
\author{T.~Patel}
\affiliation{Department of Physics, Hampton University, Hampton, VA 23668, USA}
\author{A.~Prosnyakov}
\affiliation{Racah Institute of Physics, The Hebrew University of Jerusalem, Jerusalem, Israel}
\author{R.~D.~Ransome}
\affiliation{Rutgers, The State University of New Jersey, Piscataway, NJ 08855, USA}
\author{R.~Ratvasky}
\affiliation{Department of Physics, The George Washington University, Washington, D.C. 20052, USA}
\author{H.~Reid}
\affiliation{Randall Laboratory of Physics, University of Michigan, Ann Arbor, Michigan 48109, USA}
\author{P.~E.~Reimer}
\affiliation{Physics Division, Argonne National Laboratory, Lemont, Illinois 60439, USA}
\author{G.~Ron}
\affiliation{Racah Institute of Physics, The Hebrew University of Jerusalem, Jerusalem, Israel}
\author{T.~Rostomyan}
\affiliation{Paul Scherrer Institute, Villigen CH-5232, Switzerland}
\author{O.~M.~Ruimi}
\affiliation{Racah Institute of Physics, The Hebrew University of Jerusalem, Jerusalem, Israel}
\author{K.~Salamone}
\affiliation{Center for Frontiers in Nuclear Science, Stony Brook University, Stony Brook, NY 11794, USA}
\author{N.~Sparveris}
\affiliation{Department of Physics, Temple University, Philadelphia, PA 19122, USA}
\author{S.~Strauch}
\affiliation{Department of Physics and Astronomy, University of South Carolina, Columbia, SC 29208, USA}
\author{D.~A.~Yaari}
\affiliation{Racah Institute of Physics, The Hebrew University of Jerusalem, Jerusalem, Israel}

\date{\today}

\begin{abstract}
Human bias is capable of changing the analysis of measured data sufficiently to alter the results of an experiment. It is incumbent upon modern experiments, especially those investigating quantities considered contentious in the broader community, to blind their analysis in an effort to minimize bias. The choice of a blinding model is experiment specific, but should also aim to prevent accidental release of results before an analysis is finalized. In this paper, we discuss common threats to an unbiased analysis, as well as common quantities that can be blinded in different types of nuclear physics experiments. We use the Muon Scattering Experiment as an example, and detail the blinding scheme used therein. 
\end{abstract}


\maketitle

\section{Introduction}
It is a well-established fact that biases of the analyzer of experimental data can affect the outcome~\cite{Franklin,doi:10.1146/annurev.nucl.55.090704.151521}. This is especially true in the case that strong expectations for the true value exist, for example from earlier measurements or from theoretical predictions. This is further intensified if the measured quantity is under contention, as in the case of the proton radius puzzle~\cite{naturePRP}. The way biases are introduced varies greatly, as many analyses require a large number of somewhat arbitrary choices, such as in the choice of which constraints are applied, of the exact constraint conditions, data quality selectors, fit function definition, and fit range. 

We note here that such a bias does not require the analyzer to actively seek to bias the result and can be instead attributed to decision-making with finite knowledge. A classic example is the decision that all systematic effects are accounted for. It is impossible to decide whether a systematic effect was not conceived of and therefore missed, or if such a systematic effect does not exist. If a result is close to an expected value, an analyzer (or group of analyzers) is more likely to conclude that indeed all systematic effects are included, while a result far away from the expected value will prompt further investigation.

While some techniques exist to manage and reduce biases, for example, the determination of suitable fit function and ranges from pseudo data, ideally prior to the publication of the data, as has been done in~\cite{PhysRevC.98.025204}, the most reliable way to avoid most sources of bias is ``blinding". In a blinding scheme, analyzers work with real data, but important information required to arrive at an accurate result is withheld---the analyzers are blind to them~\cite{NatureBlind}. Only after the full analyses have been developed and vetted, the data set is unblinded, and the result is revealed~\cite{PhysRevLett.126.141801}.

This technique is widespread in many physics disciplines, like atomic physics, but not often employed in scattering experiments. This is partially caused by the fact that, depending on the aimed-for physical quantity, it is hard to blind scattering experiments in a way that is not trivially defeated by the redundancy in the data but does not hinder the development of a full analysis. 

Blinding can provide, as an additional benefit, protection against a premature or unsanctioned release of results, be it from an absentminded inclusion of a result or ``almost-result" plot in a presentation or the intentional release by a collaboration member outside of a collaboration-established vetting and publication approval scheme.

In the following, we discuss the design and implementation of a blinding scheme for scattering experiments. The scheme is implemented for the Muon Scattering Experiment (MUSE) and we will use this experiment as an example, as its broad physics output covers many cases.

\section{Threat model}
The design of a suitable blinding approach depends strongly on the threat model, i.e., the scenarios the blinding scheme should defend against. 

It is very impractical to defend against an intentional, malicious actor with significant computing resources. It would be antithetical to the idea of a healthy collaboration to control the access to analysis source code, so blinding mechanisms in the software can easily be patched out. A possibility is the release of blinded-only data to the collaboration. This necessitates either the online application of a reversible blinding scheme during data taking, or an effective doubling of the data set, with an unblinded original data set under access control and a blinded version available for the collaboration. The former risks the analyzability of the data in case the unblinding scheme is found to be defective, while the latter incurs large resource consumption and requires a complete reprocessing of the data for unblinding. We also feel that a malicious actor has a multitude of other avenues of inflicting damage to the collaboration. Defense against such malicious actors is therefore not part of our threat model.

Instead, we want to counter two threats: the biasing of analyzers by the results, and the accidental premature release of results. The blinding system needs to be robust against accidental unblinding, for example by aggregation of different data subsets, or changes to the code.

Additionally, the scheme should be robust against low-level malicious attacks. An analyzer should not be able to easily unblind by just commenting out a line in the code, or by throwing a command line switch. Similarly, a produced blinded output should not be unblinded easily by a competent outsider, such as by applying a known or reconstructable correction formula.

Finally, the unblinding should be controlled by a plurality of a subset of collaboration members, or potentially by the majority of the collaboration board, and should not depend on the will or availability of a single person. 

\section{Blinded quantities}
It is worthwhile to discuss the possible physical quantities that are typically extracted from scattering experiments and how they could be blinded:

\begin{enumerate}
    \item Cross sections and radii: Cross sections are typically derived from a ratio of measured counts to predicted counts, where the latter is typically determined by Monte Carlo simulations and the calculated luminosity. Potentially, the easiest approach is to blind the true luminosity value. However, the relative change of cross sections as a function of some kinematics variable quite often actually contains the interesting physics information, so that the absolute normalization does not matter. In other cases, redundancy in the data or known cross sections in certain parts of the covered kinematic range can allow one to trivially recover the true luminosity. A prominent example is the extraction of radii, which utilizes a fit of the cross section with floating normalization and extrapolation to zero momentum transfer, where the cross sections are known. Such a fit recovers any normalization, typically to much better precision than what is known from first principles and is sensitive to the functional form of the kinemtatic dependence of the cross sections.  A blinding via the luminosity is therefore rarely viable. Instead, an effective blinding must change the measured counts and/or the predicted counts, with a hidden kinematic dependence. While a modification of the measured counts typically only allows a reduction, for example by eliminating a subset of events, the modification of predicted counts can go in both directions, for example by modifying the Monte Carlo weights.
    
    \item Cross-section ratios: In many experiments, ratios of extracted cross sections are the quantities of interest. In the case of MUSE, these include for example a charge ratio to determine two-photon exchange, $R_{2\gamma}=\frac{\sigma_{e^+p}}{\sigma_{e^-p}}$, or the species ratio $\frac{\sigma_{\mu p}}{\sigma_{ep}}$ to test lepton universality~\cite{10.21468/SciPostPhysProc.5.023}.
    Ratios of extracted cross sections can often be formed without applying Monte Carlo corrections, as many systematic effects, such as efficiencies, cancel out in the ratio, without the need for Monte Carlo normalization. To blind such ratios, the blinding not only has to affect the measured counts, but also do so in a species or charge-dependent way, as otherwise the blinding would again cancel.

    We note here that sometimes such cancellation can be desirable: a standard technique to gauge the existence and correction of local acceptance or efficiency effects in large area detectors is the comparison of two $\phi$ segments. For MUSE left vs.\ right side~\cite{gilman2017technical}, or in the case of CLAS-12 of one sector with another~\cite{BURKERT2020163419}. The underlying physics is typically $\phi$-independent and drops out in the ratio. A $\phi$-independent blinding would cancel as well and would allow a check of the acceptance. However, this check could bias the analyzer; the acceptances could be the same and give the impression the acceptance is correct. Here one has to balance the possible bias with the availability of important cross checks. In the case of MUSE, the blinding is chosen to allow for such comparisons.    
    
    \item Polarization asymmetries: Polarization asymmetries are a special case of ratios, with the addition of polarization information. For some experiments, this allows one to blind the data in a new way by hiding or modifying the beam polarization information encoded in the raw data stream. Flipping the recorded beam polarization for a subset of events will effectively dilute any asymmetry but has minimal impact on most analysis work. As the beam polarization cannot be reconstructed from detector information and must be provided by the accelerator, accidental unblinding by clever analysis is very unlikely. However, timing information might make it possible to guess the correct polarization to undo the blinding. 
    
    In the case of MUSE, where no polarization degrees of freedom are used, the charge ratio/asymmetry also depends on the external information, in the form of the magnetic field direction of the particle channel~\cite{PhysRevC.105.055201}. However, each data file has constant field, and switches are rare, making it too easy to reconstruct the proper field. The particle species, on the other hand, is reconstructed from detector timing information and is thus hard to blind. Further, it is not clear if blinding the particle ID would blind other observables than the species ratio to a satisfying degree.

    \item Coincidence measurements with two or more detectors: Experiments that measure the process in coincidence between multiple detectors can be blinded via event mixing. For example, a certain type of dark sector searches like DarkLight~\cite{Cline_2022} reconstruct the mass of the intermediate particle from the measured lepton decay pair in two spectrometers. Offsetting the event identification number, i.e., combining the information of one spectrometer from event $N$ with the information of the second spectrometer for event $M\neq N$, effectively destroys any correlation between the two spectrometers and causes any possible resonance to vanish. Subsequently, what remains is indistinguishable from background from random coincidences.

\end{enumerate}

\section{Approach}

In contrast to many other disciplines, where a single withheld calibration constant, such as, the exact frequency of a laser for spectroscopy, can effectively blind the analysis, scattering experiments need to apply a blinding function that depends on at least a subset of the same kinematical variables as the underlying physics. 

MUSE's physics goals include measurements of cross sections and ratios. This implies blinding of at least the measured data, as blinding of the simulation is not sufficient to effectively blind ratios.  For MUSE, cross sections and ratios are typically expressed as functions of beam momentum and scattering angle (or $Q^2$), and the chosen blinding function will depend on the same observables.

A scattering experiment, at its core, is a counting experiment. Blinding must then modify the counts. In principle, it is possible to change the count in both directions either by duplicating or removing events. However, duplicated events are easy to detect and therefore could be easily unblinded. The algorithm used in MUSE therefore only rejects events.

Like many experiments, MUSE models the analysis as a chain of operations on the recorded data. An optimal point in this chain has to be found to apply blinding.
In the first level of analysis, the raw data are processed to produce calibrated and normalized hits. This process typically only requires information of a given detector and does not take into account information of other detectors.
In the second layer, the hit information from multiple detectors are combined, for example for time-of-flight determinations or tracking. The individual physics analyses are based on top of these common analysis steps, defining particle- and reaction-identification selection, background suppression, high-level physical quantities etc. 

Blinding has to happen before the last level, as the decisions made for these selections are those that could be most likely affected by the bias of the analyzer. Vice versa, blinding too early might hinder the proper analysis of required calibrations on the detector level. MUSE therefore chooses to blind at the tracking step.

This choice has three benefits: the tracking step is the most time-intensive operation and therefore requires large CPU resources to redo. Blinding at the tracking level therefore protects against low-level efforts to unblind, as this would require a resource-intensive re-tracking of a large subset of the data with a patched tracker version. Further, blinding just after this step, with a suitable retention of the blinded information, allows for rapid unblinding, as only the later analysis layers need to be reprocessed. Third, the blinding code can make use of the tracker information to determine the kinematic-dependency of the blinding function.

\begin{figure}
    \includegraphics[width=\linewidth]{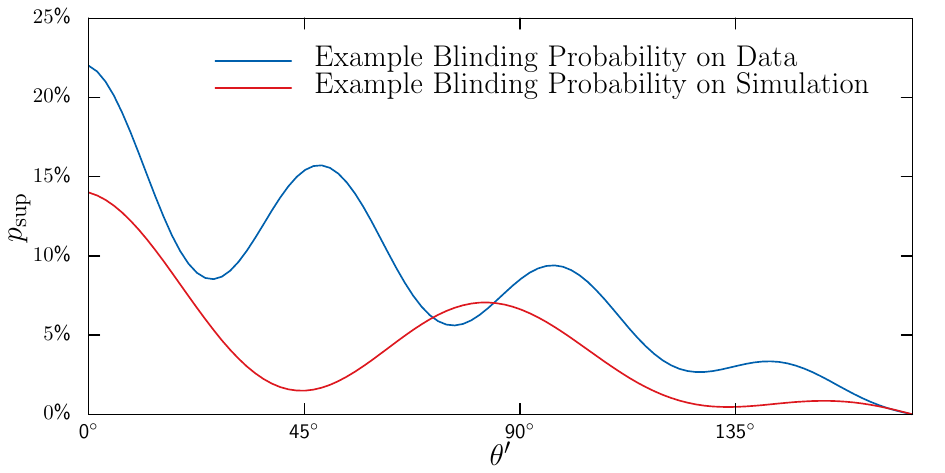}
    \caption{Two example distributions for blinding simulation or data as a function of $\theta^{\prime}$. For simulation (data) the values of $A=0.4$ ($0.8$) and $B=4.1$ ($7.2$) were arbitrarily chosen. \label{fig:exampleBlinding}}
\end{figure}

In the case of MUSE, the actual blinding is implemented as a stochastic suppression of events. 
To this end, for each track, the blinding framework calculates a probability of suppression, defined as: 
$$
p_{\text{sup}}=\frac{0.2}{3}\left(A_i+0.3\cos B_i\theta'\right)(3-\theta')
$$
with parameters $A_i \in [0.25,1]$ and $B_i \in [3,10]$, and $\theta'$ is the angle of the outgoing particle measured with respect to the nominal beam axis. The parameters $A_i$ and $B_i$ are generated from a fixed-seed pseudo-random number generator, i.e., they do not appear in clear text in the source code. In total, $2\times18=36$ sets of parameters are generated, one each for each species ($e$,$\mu$,$\pi$), charge ($\pm$), and momentum ($115, 160, 210$ MeV/$c$) combination (18 total), with a second set for simulated instead of real data. For each event, the corresponding parameter set number ($i$) is calculated from the particle ID and slow control information, then, the suppression probability, $p_{\text{sup}}$, is calculated based on the reconstructed track angle, $\theta^{\prime}$. Finally, a random number generator throws a number between 0 and 1. If the number is smaller than $p_{\text{sup}}$, the event will be suppressed. Two example probability distributions for blinding simulation and data are shown in Fig.~\ref{fig:exampleBlinding} which have for simulation (data) the values of $A=0.4$ ($0.8$) and $B=4.1$ ($7.2$).

\begin{figure}
    \centering
    \includegraphics[width=\linewidth]{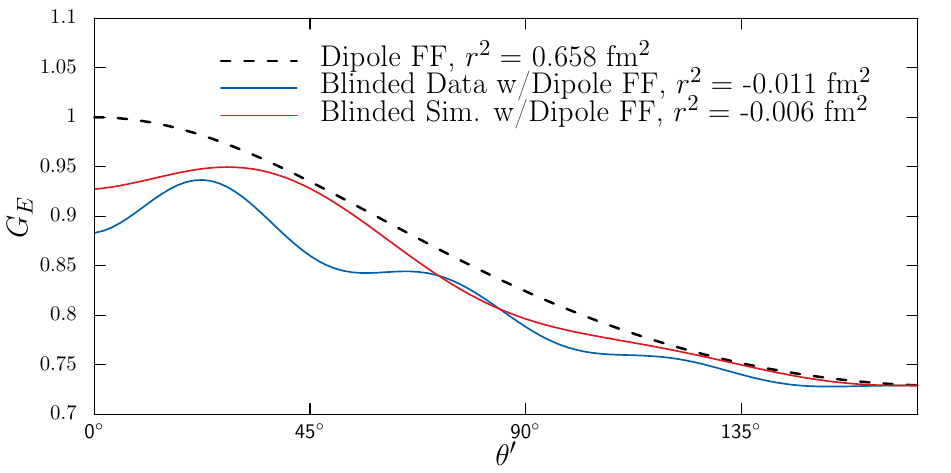}
    \caption{The impact of blinding on the extraction of the form factor. We assume that the form factor follows the standard dipole for electron-proton scattering, which is plotted for reference. The $r^2$ denotes the proton charge radius from the associated form factors. Note that the blinding directly changes the extracted radius from the true radius. To generate these form factors a momentum of 210 MeV/$c$ was used.}
    \label{fig:blindedGE}
\end{figure}

If we blind on the cross-section level, then the impact of blinding on the form-factor level follows the square-root of the blinding probability. An example of the change to the extracted form factor in simulation and data can be seen in Fig.~\ref{fig:blindedGE}, where we assume a dipole form factor as the extracted form factor; also shown is the extracted $r^2$, the proton charge radius squared, for the dipole form-factor, and for the simulation and data with blinding factors applied.

\begin{figure}
    \centering
    \includegraphics[width=\linewidth]{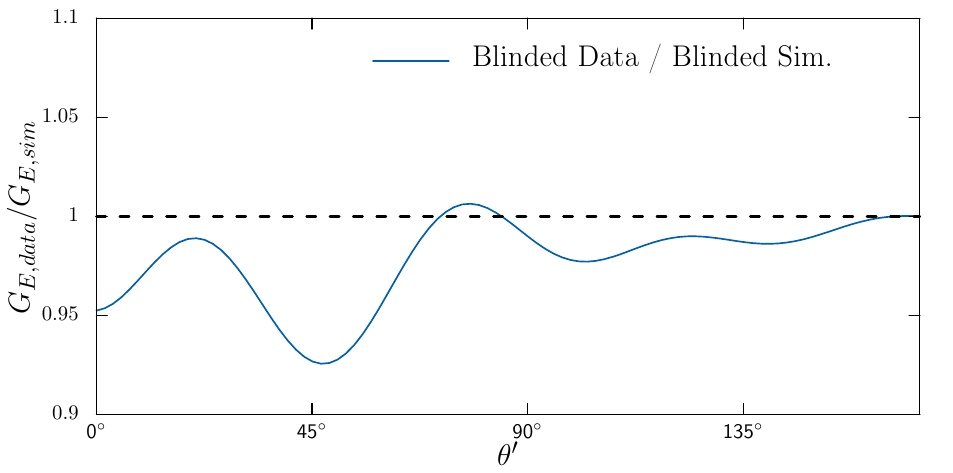}
    \caption{The ratio of data to simulation of the extracted form-factor, assuming a dipole shape. A line at 1 is shown to guide the eye.}
    \label{fig:ratio}
\end{figure}

In the final analysis, one would divide the data by the simulation to correct for detector acceptance, energy loss, target thickness, etc. It is worth emphasizing that with different blinding factors applied to the data and simulation, an overall blinding effect is preserved when taking ratios of data and simulation. An example of the ratio of form factors from simulation and experiment is shown in Fig.~\ref{fig:ratio}.

\section{Cryptographic blinding}

A naive implementation of blinding could then simply remove the selected track from the output file. However, this would require a full re-tracking of the data to unblind. Instead, the code marks the event as blinded. Further analysis steps written using the MUSE analysis framework will automatically skip these events if unblinding is not enabled. 

Of course, such blinding would be trivially defeated by simply ignoring the flag in the following steps. Therefore, the code encrypts the track information. At the beginning of the processing of a data file, for each $i$, a 32-bit random number $X_i$ is generated. The track information is then encrypted by XORing the bit-representation of the track parameters with $X_i$ for the selected key slot, repeating the 32-bit pattern as necessary to cover all bits of the track information. 
We note here that this is not cryptographically secure. The XOR key can most likely be reconstructed from the encrypted information and known or guessable bits of the cleartext. However, as discussed above, such an attempt is outside of our threat model.

\section{Unblinding schedule and logistics} 

To facilitate unblinding, the per-file random keys are stored in the output file, however, encrypted using a public/private key system. MUSE makes use of the GPG cryptographic library~\cite{GPGLibrary} for this purpose. 

In a public/private key system, one direction of the encryption/decryption process is performed with a public key, that is, the key is assumed to be known to the world~\cite{GPGHandbook}. The other direction of the encryption/decryption then requires the private key, supposed to be known only by the respective key holders. 

It is desirable that the decision and capability to unblind is not in the hands of a single individual. To enforce this algorithmically, MUSE identified three persons to hold keys. The algorithm requires two out of the three keys to unblind, in turn requiring two out of three of the key holders to agree. This avoids a single point of failure and a single point of trust scenario.

The framework holds all public keys to automatically encrypt the $X_i$. To do so, every $X_i$ is encrypted three times, with the three 2-out-of-3 combinations of the public keys. Any combination of two private keys can then be used to unlock one of the encrypted versions to decrypt $X_i$ and unblind the file. This is done automatically by the analysis framework when the private keys are supplied.

Blinded analysis has the risk that problems in the data are hidden by the blinding, and a relevant systematic effect is only discovered after unblinding. In this case, most likely, the data would not be abandoned, but the required corrections would be applied. For this scenario, the analysis has lost some of the crucial benefits of blinding, as now bias can influence the results. Even so, the blinding retains the protection against accidental publication before the unblinding, and the bias can be mitigated by policy, for example by freezing the analysis except for the required changes. 

The danger of such an occurrence can be reduced if data sets can be unblinded selectively.

The implementation for MUSE allows for different keys for each $i$. Unblinding can then be species, charge, and momentum-specific. Further, it is planned to change the keys (and seeds for the random number generators) for different data-taking periods. This allows to unblind data subsets. 

The current plan for MUSE aims to unblind the pion-proton scattering data first. This allows validation of the analysis with data that are not the focus of debate. Any deficiencies discovered in such analysis would then be corrected before further data sets are unblinded. 

It is then planned to unblind the remaining data in accordance with the availability of finalized analyses for the corresponding quantity. MUSE expects that two-photon exchange studies from the ratio of electron to positron scattering come first, followed by the electron-to-muon universality ratio, with the combined analysis and extraction of the proton radius from the whole data set coming last.

\section{Conclusion}
In this paper, we discussed the applicability of blinding to precision scattering experiments, with MUSE as a  case study. Our threat model and blinding scheme does not include intentional malicious actors, but instead defends against accidental release of results and, most importantly, explicit and implicit biases introduced in the analysis. Built with standard cryptographic building blocks, the scheme implemented for MUSE enables a fine-grained, stepwise unblinding schedule conditioned on a two-of-three consensus, avoiding all single points of failure. We believe this approach is easily transferable to other scattering experiments and analysis codes.
\vspace{1.5em}
\section{Acknowledgments}
We acknowledge and thank the Paul Scherrer Institute for its hospitality and support.
This work was supported by the US National Science Foundation (NSF) grants 1436680, 1505934, 1614456, 1614773, 1614850, 1614938, 1649873, 1649909, 1807338, 1812382, 1812402, 1913653, 2012114, 2113436, and 2012940 by the United States–Israel Binational Science Foundation (BSF) grant 2012032, 2017673, and 2017630, by the U.S. Department of Energy (DOE) with contract no. DE-AC02-06CH11357, DE-SC0012589, DE-SC0016577, DE-SC0012485, DE-SC0016583, DE-SC0019768, DE-SC0014448, and DE-FG02-94ER40818, and by PSI, Villigen, Switzerland.

\bibliography{blindingbib}

\end{document}